\documentclass [preprint,showpacs,preprintnumbers,amsmath,amssymb, nofootinbib]{revtex4-1}
\usepackage{graphicx}  
\usepackage{amsmath,amssymb}   
\usepackage{color}
\definecolor{darkgreen}{rgb}{0,0.35,0}

\newcommand{\be}{\begin{equation}}
\newcommand{\ee}{\end{equation}}
\newcommand{\bea}{\begin{eqnarray}}
\newcommand{\eea}{\end{eqnarray}}
\newcommand{\bref}[1]{(\ref{#1})}
\newcommand{\nn}{\nonumber}
\newcommand{\A}{\alpha}

          \newcommand{\w}{\omega}
          
\newcommand{\h}{\eta}           
           
\newcommand{\W}{\Omega}

\newcommand{\sech}{{\rm sech}}

\def\6{\partial} \def\7{\tilde} \def\8{\hat}

\def\d{\dot}

\def\pa{\partial}

\def\CL{{\cal L}}

\def\vs{\vskip 3mm}\def\={{\;=\;}}\def\+{{\;+\;}}

\def\lag{Lagrangian}
\def\ebox#1#2{\vskip 2mm{\vbox{\hrule\hbox{\vrule\kern3pt\vbox{\kern3pt
         {\begin{eqnarray}#1\label{#2}\end{eqnarray}}
         \kern3pt}\kern3pt\vrule}\hrule}}\vskip 2mm}
\def\tbox{\vskip 2mm{\vbox{\hrule\hbox{\vrule\kern3pt\vbox{\kern3pt
         {{\hfill {\small ${}^{notebook\; kiyoshi
         }$} \\
         \large \bf ~~\reptitle}\\ } 
         \kern3pt}\kern3pt\vrule}\hrule}}\vskip 2mm}
\def\lag{Lagrangian }\def\lags{Lagrangians }
\def\vs{\vskip 4mm}

\def\dif{{\rm d}}
\def\deriv{\@ifnextchar[{\@deriv}{\@deriv[]}}
   \def\@deriv[#1]#2#3{\mathchoice%
{{\dif^{#1}#2\over\dif{#3}^{#1}}}{{\dif^{#1}#2/\dif{#3}^{#1}}}%
{{\dif^{#1}#2\over\dif{#3}^{#1}}}{{\dif^{#1}#2/\dif{#3}^{#1}}}}

\usepackage{amssymb}
\usepackage{amsmath}
\usepackage{latexsym}
\usepackage{color}

\def\d2derpar#1#2{\mathchoice%
{{\partial^2 #1\over\partial #2^2}}%
{{\partial^2 #1/\partial #2^2}}%
{{\partial^2 #1\over\partial #2^2}}%
{{\partial^2 #1/\partial #2^2}}%
}

\def\dif{{\rm d}}
\def\deriv{\@ifnextchar[{\@deriv}{\@deriv[]}}
   \def\@deriv[#1]#2#3{\mathchoice%
{{\dif^{#1}#2\over\dif{#3}^{#1}}}{{\dif^{#1}#2/\dif{#3}^{#1}}}%
{{\dif^{#1}#2\over\dif{#3}^{#1}}}{{\dif^{#1}#2/\dif{#3}^{#1}}}}

\begin{document}
\preprint{ICCUB-14-053}
\title{{Dynamical sectors for a spinning} particle in AdS$_3$}
\author{Carles Batlle$^1$}\email{carles.batlle@upc.edu} \author{Joaquim Gomis$^2$}\email{gomis@ecm.ub.es} \author{Kiyoshi Kamimura$^3$}\email{kamimura@ph.sci.toho-u.ac.jp} \author{Jorge Zanelli$^4$}\email{z@cecs.cl}
\affiliation{ $^1$Universitat Polit\`ecnica de Catalunya-BarcelonaTech,  Spain.}
\affiliation { $^2$Departament d'Estructura i Constituents de la Mat\`eria and Institut de Ci\`encies del Cosmos, Universitat de Barcelona, Diagonal 647, 08028 Barcelona, Spain}
\affiliation{$^3$Department of Physics, Toho University, Miyama, Funabashi, 274-8510, Japan}
\affiliation{$^4$Centro de Estudios Cient\'{\i}ficos (CECS) Casilla 1469 Valdivia, Chile and Universidad Andr\'es Bello, Rep\'ublica 440, Santiago, Chile}
\date{\today}
\begin{abstract}Abstract:
We  consider the dynamics of a particle of mass $M$ and spin $J$ in AdS$_3$. The study reveals the presence of different dynamical sectors depending on the relative values of  $M$,  $J$ and the AdS$_3$ radius $R$. For the subcritical  $M^2 R^2-J^2 >0$  and supercritical  $M^2 R^2-J^2<0$ cases, it is seen that the equations of motion  give the geodesics of AdS$_3$. For the critical case  $M^2R^2=J^2$ there exist  extra gauge transformations which further reduce the physical degrees of freedom,  and the motion corresponds to the geodesics of AdS$_2$. This result   should be useful in the holographic interpretation of the entanglement entropy for 2d conformal field theories with gravitational anomalies.
\end{abstract}

\maketitle


\vs
\section{Motivation and results}      

Point particles in  spacetime  offer a simplified setting to study the interplay between the geometry of spacetime and quantum mechanics.  As noted long ago by Deser, Jackiw and 't Hooft, point particles in $2+1$ gravity are naked conical singularities in which the mass is related to the angular deficit. The energy-momentum tensor has  Dirac deltas with support at the position of the particles, which induces delta singularities in the curvature at those points. Infinite curvature concentrated at an isolated point corresponds to a conical singularity produced by the removal of a wedge. These conical defects do not affect the local geometry on open sets that do not include the singularities, but they change the global topology \cite{DJtH}. 

Point particles in AdS$_3$ are also related to black holes. Since point particles are conical singularities, they are obtained by identification in AdS$_3$ by a global spacelike Killing vector with a fixed point in a similar way as the 2+1 black hole is obtained by identifications in the universal covering of AdS$_3$ \cite{BHTZ}. The spacetime geometry of a conical singularity in AdS$_3$ is identical to the BTZ geometry, where the mass of the black hole is minus the mass of the point particle, and therefore a point particle of mass $M$ in 2+1 can be viewed as a black hole of mass $-M$ \cite{MZ}.

\begin{figure}
\begin{center}
\includegraphics[scale=0.5]{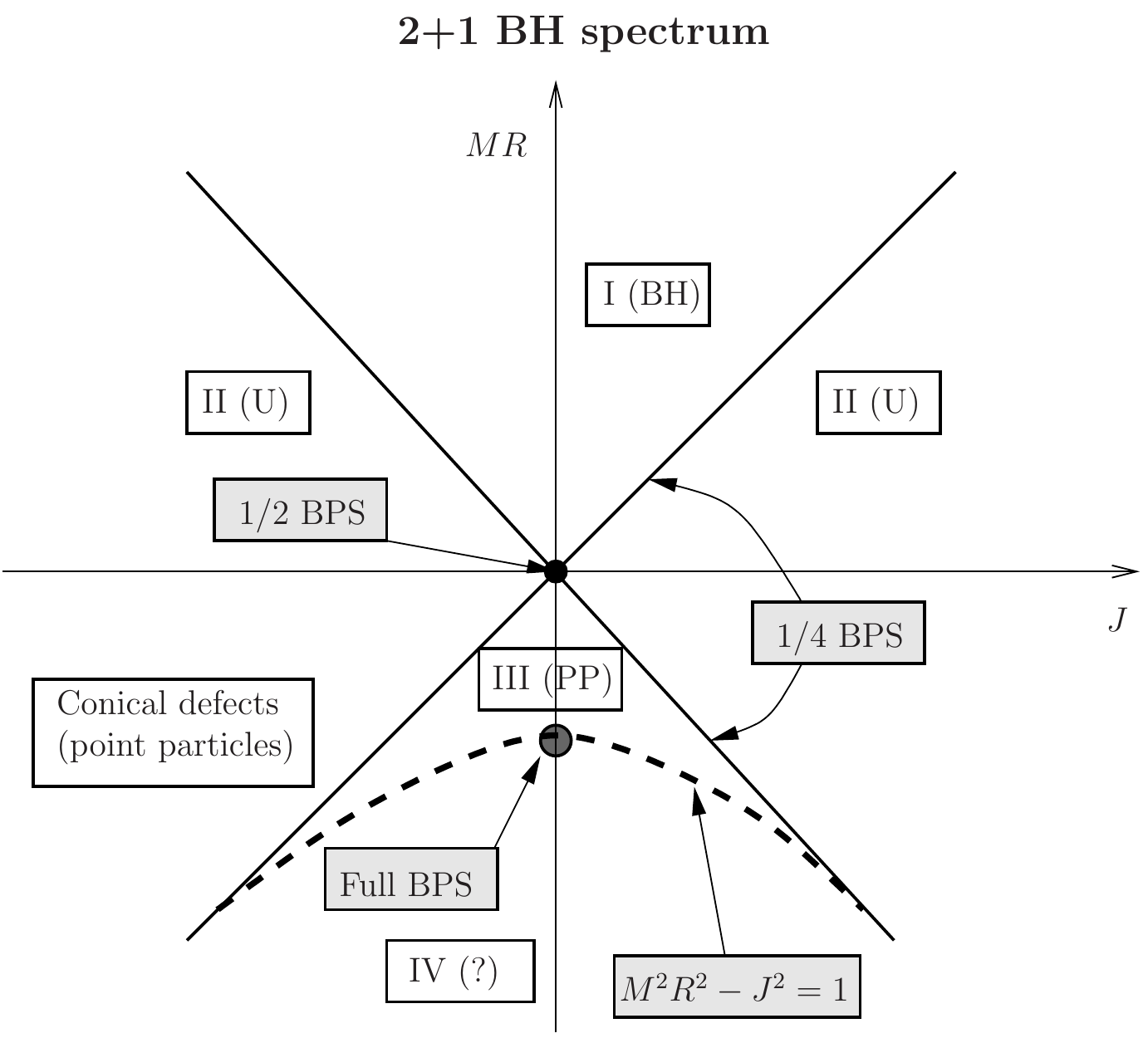}
\end{center}
\caption{$J$-$M$ plot for locally AdS$_3$ geometries: Physical black holes ($MR\geq |J|$), point particles ($MR \leq -|J|$) and unphysical states ($|M|R \leq |J|$). The critical cases $|M|R=\pm |J|$ correspond to extremal configurations.}
\label{figbhspectrum}
\end{figure}

The 2+1 black hole can have angular momentum $J$ and by the same token, a point particle can be endowed with spin. {An important quantity that characterizes both black holes and point particle states is $ \kappa \equiv J^2-M^2 R^2$, where $R$ is the AdS radius. Non-extremal black holes and spinning  point particles correspond to $\kappa <0$ (subcritical case). The extreme (critical)} geometries, $\kappa=0$, have additional special features, like admitting globally defined Killing spinors (supersymmetric BPS states). Figure \ref{figbhspectrum} displays the different sectors of 2+1 BH-particle states in the $M$-$J$ plane. The AdS$_3$ geometry, without identifications, is the point $J=0, M R=-1$; black holes cover region I ($\kappa \leq 0, M>0$); point particles are described by III ($\kappa \leq 0, M<0$). The supercritical regions II ($\kappa>0$) are unphysical states. States below the hyperbola $\kappa=-1, M<0$ correspond to angular excesses rather than defects, and also display delta-like curvature singularities at $r=0$.
On the other hand, it is also possible to think of a particle as a probe that does not significantly affect the spacetime geometry around it, a localized perturbation with negligible back reaction. The appropriate setting to describe this would be an action for a relativistic point particle whose trajectories, in the absence of other interactions, are geodesics of the spacetime background. In that case $M$ and $J$ are parameters in the action and  $\kappa$ is an intrinsic feature, while for the  black hole they are constants of motion that depend on the initial conditions.

The geodesic equation is a local statement and is therefore the same for a particle on a patch of AdS$_3$ or around a 2+1 black hole.  The only difference would be in the orbits, since they depend on the global properties of the manifold. Hence, the conserved quantities for the different orbits could be the same, \textit{e.g.} energy and angular momentum, but their specific values would determine the class of geodesics that the particle traces. 

Motion of particles in AdS$_3$ can  also be related to the holographic entanglement entropy. As shown in \cite{Ryu:2006bv}, the entanglement entropy in AdS$_D$ is {related to minimal $(D-2)$-surfaces} in AdS$_D$. For the case of AdS$_3$, the minimal surface is the length of the  geodesics, and these are the trajectories of particles in AdS$_3$.  The holographic description of the entanglement entropy for conformal field theories in two dimensions with gravitational anomalies \cite{AlvarezGaume:1983ig} has been studied in \cite{Castro:2014tta}.
 
Motivated by these observations, we consider the dynamics of a particle of mass $M$ and spin $J$ in AdS$_3$. The \lag equations of motion reveals the presence of different dynamical sectors depending on $M, J$ and $R$. For the subcritical  ($\kappa <0$) and supercritical ($\kappa >0$) cases it is seen that the equations of motion  give the corresponding geodesics of AdS$_3$. For the critical case  $(\kappa=0)$ there exists an extra gauge transformation which further reduces the physical degrees of freedom.\footnote{This gauge transformation is the bosonic analog of kappa symmetry \cite{deAzcarraga:1982dw} \cite{Siegel:1983hh}  for superparticles \cite{Casalbuoni:1976tz} \cite{Brink:1981nb}  that kills half of the degrees of freedom.} The orbits correspond to geodesics of AdS$_2$.

The presence of dynamical sectors appears also in a (2+1)D harmonic oscillator system with exotic Newton-Hooke symmetry. The system displays three different phases depending on the values of the parameters \cite{Alvarez:2007fw}. The reduced phase space description reveals a symplectic structure similar to that of Landau problem in the non-commutative plane \cite{DH1,HP1}. There is a close relation between the (2+1)D exotic Newton-Hooke symmetry and the non-commutative Landau problem \cite{Alvarez:2007ys}.\footnote{The motion of an anyon in an electromagnetic field also has sectors, see e.g.\cite{PHMP}.}

The non-relativistic limit of a  spinning particle in  AdS$_3$ shows the Newton-Hooke symmetry and gives the (2+1)D  exotic harmonic oscillator system.

Since the AdS$_3$ algebra can be written as a sum of two chiral $so(2,1)$ factors, we construct a Lagragian for spinning particles in terms of chiral coordinates. The relation between the chiral and non-chiral variables is given in terms of a differential equation that can be solved perturbatively. The equations of motion for the critical case in the chiral formulation describe orbits that are geodesics of AdS$_2$.

The subcritical case $(\kappa<0)$ includes a special representative, $J=0$, while $M=0$ is a representative for the supercritical case $(\kappa>0)$. In each case we explicitly give the coordinate transformation that takes the \lag into one described by the corresponding representative. We expect the points  $J=0$ or  $M=0$ to be described in terms of non-commutative coordinates, in a similar way as for the flat case (see \textit{e.g.} \cite{Skagerstam:1989ti}).

Some of the particle states have a lowest energy (BPS) bound. In particular, the critical sector with $\kappa=0$, with $M>0$ saturates a BPS bound and therefore is a candidate to be a supersymmetric configuration, with $1/4$ SUSY, if the system is embedded in a supersymmetric model. These particles belong to the BH sector of the $J$-$M$ plane. The subcritical configurations with $M<0$ do not have a lowest energy bound but an upper one, and therefore do not seem to correspond to stable BPS states. As shown in \cite{MZ}, the critical states with $M<0$ also admit globally defined Killing spinors. Nevertheless, a complete correspondence with the BH spectrum should not be expected since we assume the particle as a probe that does not modify the spacetime background.

Summing up, we prove the existence of sectors in the dynamics of a spinning particle in AdS$_3$. The critical sector has half the physical degrees of freedom due to existence of an extra, kappa-like bosonic gauge symmetry. The presence of sectors is in correspondence with the spectrum of BH. Our results shed light on the  holographic interpretation of the entanglement entropy for CFT$_2$ with gravitational anomalies \cite{Castro:2014tta}, for which  the left $c_L$ and right $c_R$ central charges are different, and one will have sectors for these class of theories. The critical sector corresponds to chiral CFTs with only left or right moving sector, while the sign of $c_L c_R$ determines the  subcritical ($+$) and supercritical ($-$) sectors. Any particle with spin $J$  in a locally  AdS$_3$ metric will exhibit the same kind of dynamical sectors, but they will be absent, for instance, for  local dS$_3$  geometries. The analysis should also be useful in the study of the  motion of anyons around a BTZ black hole.
 
The paper is organized as follows. Section \ref{secL} presents the Lagrangian, discusses the dynamical sectors and the equations of motion. Section \ref{secChiral} introduces the chiral variables and analyzes the dynamical sectors in this description. The relation between chiral and nonchiral variables is established in Section \ref{secNch2ch} by means of a set of nonlinear differential equations that can be solved perturbatively.  Section \ref{secTrans} presents an explicit biparametric  transformation, in terms of the chiral variables, that connects Lagrangians with different $M$, $J$ parameters in the super- and subcritical sectors. Finally, Section \ref{secBPS} uses the chiral form of the Lagrangian to discuss the existence of BPS energy bounds for several values of the parameters.

\section{AdS$_3$ action and equations of motion}\label{secL}    

The action of a massive spinning particle in AdS$_3$ is constructed from the coordinates of the world line  $x^a(\tau)$ and a generic Lorentz transformation $\Phi_a{}^b(\tau)$.\footnote{ The tangent space metric is $\h_{ab}=diag (-;++)$. Indices $m, n, \ldots$ refer to  the spacetime manifold where the particle moves. $a,b,\ldots=0,1,2$ are tangent space indices, and $a',b',\ldots=1,2$.} To lowest order in derivatives, the action is \cite{Skagerstam:1989ti,deSousaGerbert:1990yp}
\begin{eqnarray} \label{Action02}
&& I_0[x,\Phi] =-M \int d\tau\, (\dot{x}^m e^a{}_m \Phi_a{}^0)\;\nn\\
&&-\frac{J}{2}\int d\tau \epsilon_{a'b'}\eta^{cd} {\Phi_d}^{a'} \left[{\dot{\Phi}_c}{}^{b'} + \dot{x}^m{\omega_{mc}}^e{\Phi_e}^{b'}\right],
\end{eqnarray}
where $e^a{}_m(x)$ and $\omega^{ab}_m(x)$ are the dreibein and spin connection of AdS$_3$. The \lag is given by the Maurer-Cartan (MC) form (non-linear realization \cite{Coleman}) for the coset $G/H={SO(2,2)}/{SO(2)}$. The AdS$_3$ generators verify the so(2,2) algebra
\bea
\left[P_a,P_b\right]&=&-i {R^{-2}}M_{ab},\;\left[P_{a},M_{cd}\right]=-i\h_{a[c}P_{d]}, \nn\\ 
\left[M_{ab},M_{cd}\right]&=&-i\h_{b[c}M_{ad]}+i\h_{a[c}M_{bd]}\, ,
\eea
and the stability group $H$ is generated by $M_{12}$. We consider a local parametrization of the coset element
\bea\label{nc-cosetg}
g&=&g_0\,U\in G/H,\quad g_0=e^{iP_0x^0}e^{iP_{1}x^{1}}e^{iP_{2}x^{2}}, \nn\\ 
h&=&e^{iM_{12}\A}\in H, \qquad U=e^{iM_{02}v^{1}}e^{-iM_{01}v^{2}}\, .
\eea
The Lorentz transformation $U$ can be expressed as
\be\nn
 \Phi_a{}^b=\!
\begin{pmatrix} \cosh v^1& 0&-\sinh v^1\cr 0 & 1& 0 \cr -\sinh v^1&0 &\cosh v^1 \end{pmatrix}\! \!\!
\begin{pmatrix} \cosh v^2&\sinh v^2& 0 \cr \sinh v^2 &\cosh v^2& 0 \cr 0 & 0 &1 \end{pmatrix} .
\ee
The dreibein and spin connection in this parametrization are obtained from the MC form $\Omega_{0}=-ig_0^{-1}dg_0=P_ae^a+\frac12M_{ab}\w^{ab}$,
\bea
e^0&=&
\cosh \hat{x}^1\,\cosh\hat{x}^2\,d x^0, \quad e^1=\cosh\hat{x}^2\,d x^1, \nn\\ 
e^2&=& d x^2, \qquad \w^{02}=\cosh\hat{x}^1\,\sinh\hat{x}^2\,d \hat{x}^0, \nn\\
\w^{12}&=&\sinh\hat{x}^2\,d \hat{x}^1,\quad \w^{01}=\sinh\hat{x}^1\,d \hat{x}^0,
\eea
where $\hat{x}^a=x^a/R$. The MC form associated to  $g$ is $\W=-ig^{-1}dg=P_aL^a+\frac12M_{ab}L^{ab}$,  
where
\be\label{MCnonchiral}
L^a=e^b {\Phi_b}^a,\qquad L^{ab}={\Phi_c}^a(\h^{cd}\,d+\w^{cd}) {\Phi_d}^b.
\ee
The \lag \bref{Action02}  is constructed from the pullback of  the (pseudo) invariant forms $L^0$ and $L^{12}$ as
\be\label{nonchiral}
\CL^{non}=-ML^0-JL^{12}.
\ee
The EL equations of motion can be written as 
\bea\label{homogeneous}
 && \left({-MR^2L^{01}}+{J}L^2\right)= \left({-MR^2L^{02}}-{J}L^1\right)=0,  \nn\\
 && \left({ML^{2}}-{J}L^{01}\right)=\left({ML^{1}}+{J}L^{02}\right)=0 .
 \eea
If $J^2\neq M^2R^2$ these equations become $L^{a'}=0$ and $L^{0a'}=0$. They  relate the spin variables $v^{a'}$ to the coordinates of 
 the world line by
 \be
{{\Phi}_a}^0=\frac{\dot x^m{e_m}^b\h_{ba}}{\sqrt{-g}}, \quad g\equiv\dot x^m \dot x^n g_{mn},
 \label{noncritical}
 \ee
 and  yield also the geodesic equation,
\be \frac{d}{d\tau}\frac{\dot x^m}{\sqrt{-g}} +\Gamma_{rn}^m\frac{\dot x^r\dot x^n}{\sqrt{-g}}=0 \label{EOMAdS}
\ee
for the metric  $g_{mn}={e_m}^b\h_{ba}{e_n}^a$. Note that (\ref{EOMAdS}) has no contribution from the spin variables $v^{a'}$ because the physical states in configuration space are given by $x^{a'}$ only and the $v^{a'}$ are not independent local degrees of freedom,  as can be seen from (\ref{noncritical}) as well as from the Hamiltonian analysis. In the reduced phase space, the velocity $\dot x^m$ is proportional to the momentum, and therefore the coordinates do not exhibit zitterbewegung.

In the critical case  $J^2=M^2R^2$, only two of  the equations in  \bref{homogeneous} are independent and there are new gauge symmetries, besides diffeomorphisms, that reduce the number of  degrees of freedom from 4 to 2. The sectors are also present in local AdS$_3$ or warped AdS$_3$,  but not for dS$_3$.

\section{Chiral formulation}\label{secChiral}    
 The chiral form of \bref{Action02} is obtained by making use of the isomorphism $SO(2,2) = SO(2,1)\times SO(2,1)$, namely
\be \label{Lchiral}
\CL^{ch}=\mu^+L^0_++\mu^-L^0_-,
\ee
with
\be\label{mu_MBR}
\mu^\pm=-\frac12(J\pm MR),
\ee
and  where 
\begin{eqnarray}
L_\pm^a&=&\frac12\epsilon^{abc}L_{bc}\pm L^a/R,\label{Lpm}\\
j^\pm_a &=&\frac12(-\frac12\epsilon_{abc}M^{bc}\pm RP_a),\label{jpm}
\end{eqnarray}
are the chiral MC forms and generators, related to those of AdS$_3$.  

The sub- and super-critical sectors  are defined by the sign of  $4\mu^+\mu^- = (J^2-M^2R^2)=\kappa$, and in the critical sector  either $\mu^+=0$ or $\mu^-=0$, which corresponds to an AdS$_2$ \lag in each case. The \lag \bref{nonchiral} was obtained from the  coset ${SO(2,2)}/{SO(2)}$. In order to construct  a \lag  in terms of chiral variables we would like to know in which form an element of  ${SO(2,2)}/{SO(2)}$  can be written as a product of chiral coset elements ${SO(2,1)}/{SO(2)}$. To answer  this, let us introduce
 \bea
g^+&=& (e^{ij^+_0x^0}e^{ij^+_{1}x^{1}}e^{ij^+_{2}x^{2}}e^{ij^+_{1}v^{1}}e^{ij^+_{2}v^{2}})\nn\\
&=&e^{ij^+_0X^{0}}e^{ij^+_{1}X_+^1 }e^{ij^+_{2}X_+^2}\,e^{ij^+_0\A}\nn\\
&\equiv&\7g^+\,e^{ij^+_0\A},
\nn\\
g^-&=&(
e^{-ij_0^-x^0}e^{-ij_{1}^-x^{1}}e^{-ij_{2}^-x^{2}}
e^{ij_{1}^-v^{1}}e^{ij_{2}^-v^{2}})\nn\\
&=&e^{-ij^-_0X^{0}}e^{-ij^-_{1}X_-^1}e^{-ij^-_{2}X_-^2}e^{ij^-_0\A}\nn\\
&\equiv&\7g^-\,e^{ij^-_0\A},
\label{42gminush}\eea
where the terms containing $\A$  are compensating elements of the chiral  $H=SO(2)$ factors. The elements of ${SO(2,2)}$ are then given by
\bea
g^+g^-&=&\7g^+\7g^-\,h,
\eea
where
\begin{equation}
\label{hpm}
 h=e^{ij^+_0\A}e^{ij^-_0\A}=e^{i(j^+_0+j^-_0)\A}\in H,
\end{equation}
and $(\7g^+\7g^-)\in G/H$ is parametrized by 5 coordinates $(X^0,X^{\pm1},X^{\pm2})$. Using the expressions for  $\tilde{g}^\pm$, the \lag \bref{Lchiral} can be written  as
\bea
&&\CL^{ch} =\mu^+(\cosh X_+^2\cosh X_+^1 \dot X^0+\sinh X_+^2 \dot X_+^1 )\nn\\
&&+\mu^-(-\cosh X_-^2\cosh X_-^1 \dot X^0+\sinh X_-^2\dot X_-^1),
\label{Lchiral2}\eea
and the chiral equations of motion $L_\pm^{a'}=0$, are 
\bea
&&{\dot X_\pm^{1}}(\tau) ~=\mp\tanh X_\pm^{2}(\tau) \cosh X_\pm^{1}(\tau) {\dot X^0}(\tau), \nn\\
&&{\dot X_\pm^{2}}(\tau) ~=\pm\sinh X_\pm^{1}(\tau) {\dot X^0}(\tau) .
\label{ChMCp4}\eea
In the critical case $(\mu^-=0)$, the equations of motion are $L_+^{a'}=0$  and give the geodesics in AdS$_2$, while the  $X^{a'}_-$'s are gauge degrees of freedom since they are absent from the \lag.

In the non-relativistic limit $X_{\pm}^{a'}\to  X_{\pm}^{a'}{/\omega},\,\mu^{\pm}\to\omega^2 \mu^{\pm}$, the \lag takes the Newton-Hooke form \cite{Alvarez:2007fw} up to a divergent total derivative, and the equations of motion \bref{ChMCp4} become those of two harmonic oscillators.

\section{Non-chiral to chiral variables}\label{secNch2ch}  

The relation between chiral and non-chiral coordinates is established by comparing the corresponding expressions for the coset element in the two parametrizations. Introducing an auxiliary parameter $t$ to rescale $v^i$ in the coset expressions (\ref{nc-cosetg}) and (\ref{42gminush}), one finds
\begin{eqnarray}
 \lefteqn{e^{\pm ij_0 x^0}e^{\pm ij^\pm_1 x^1}e^{\pm ij^\pm_2 x^2} e^{itj^\pm_1 v^1}e^{it j^\pm_2 v^2}}\nonumber\\
&=& e^{\pm ij^\pm _0X^0 (t)}e^{\pm ij^\pm_1 X^{\pm 1}(t)}e^{\pm ij^\pm _2 X^{\pm2}(t)}e^{ij^\pm _0\A(t)}.
\label{42gplus41}
\end{eqnarray}
Expanding in powers of $t$ corresponds to expansions in $v^i$, and differentiating with respect to $t$, yields a set of nonlinear  differential equations relating the non-chiral variables ($t=0$) and chiral variables ($t=1$), 
\bea
\lefteqn{\pm\cosh X^{\pm 2}\cosh X^{\pm1}\pa_t X^0+\sinh X^{\pm2}\pa_t X^{\pm1}+\pa_t\A}
\nn\\
&&=\sinh(tv^2)v^{1},
\nn\\
\lefteqn{\cos\A \left(\sinh X^{\pm2} \cosh X^{\pm1}\pa_t X^0\pm\cosh X^{\pm2}\pa_t X^{\pm1}\right)}
\nn\\&-&\sin\A \left(-\sinh X^{\pm1}\pa_t X^0\pm\pa_t X^{\pm2}\right)=\cosh(tv^2)v^{1},
\nn\\
\lefteqn{\sin\A \left(\sinh X^{\pm2} \cosh X^{\pm1}\pa_t X^0\pm\cosh X^{\pm2}\pa_t X^{\pm1}\right)}
\nn\\&+&\cos\A \left(-\sinh X^{\pm1}\pa_t X^0\pm\pa_t X^{\pm2}\right)=v^2.
\label{44ChMCp2}\eea
These equations can be solved as a series in $t$ with initial conditions $X^0(0)=x^0/R$, $X^{\pm1}(0)=x_1/R$, $X^{\pm2}(0)=x_2/R$ and $\A(0)=0$. To lowest order, the solution is given by
\begin{eqnarray} \label{nch2ch}
X^0(t)&=&\hat x^0-t \,v_1\ \sech\,\hat x_1 \sinh \hat x_2+O(t^3),\\
X^{\pm1}(t)&=&\hat x_1\pm t \,v_1 \cosh \hat x_2+ O(t^2),\nn\\ 
X^{\pm2}(t)&=&\hat x_2\pm t (v_2 - \,v_1\tanh \hat x_1 \sinh \hat x_2) + O(t^2) ,\nn\\
{\A}(t)&=& O(t^2).
\end{eqnarray}

\section{Transformation of \lags} \label{secTrans}    
\lags with different values of  $(M,J)$ can be related by a biparametric family of point transformations, $(X^{\pm 1},X^{\pm 2})\to (X^{\pm 1}(s_\pm),X^{\pm 2}(s_\pm))$ given by
\bea
\lefteqn{ X^{\pm1}({s^\pm})=}\nn \\ & & \cosh^{-1}\left( e^{{s^\pm}}\frac{\cosh X^{\pm1}}{\sqrt{1
+(e^{2{s^\pm}}-1)\alpha(X^\pm)}}\right)
\nn\\
\lefteqn{X^{\pm2}({s^\pm})=}\nn \\&&\cosh^{-1}\left(\cosh X^{\pm2}\sqrt{{1+(e^{2{s^\pm}}-1)\alpha(X^\pm)}}\right),
\label{infDX+123}\eea
where 
\[
\alpha(X^\pm)=\frac{\sinh^2 X^{\pm2}}{\cosh^2X^{\pm2}-\sech^2X^{\pm1}}. 
\]
The left and right chiral \lags transform as $L^0_\pm \to e^{s^\pm} L^0_\pm$, up to total derivatives.  From the relation between the chiral and nonchiral forms of the Lagragian given by (\ref{Lchiral}) and (\ref{mu_MBR}) one can change the coefficients of the $L_P$ and $L_J$ terms of the nonchiral form as 
\bea
M&\to&M'={e^{s_1} (M\cosh{s_2} + J/R \sinh{s_2})}, \nn\\
J&\to&J'=e^{s_1} (J \cosh{s_2} + MR\sinh{s_2}),
\label{MBs}
\eea
with $s_\pm = s_1 \pm s_2$, so that $\kappa \to \kappa' =e^{2s_1}\kappa$. In particular, for the subcritical sector, taking $\tanh s_2 = -J/(MR)$ makes $J'=0$, while in the supercritical case, $\tanh s_2 = -MR/J$ yields $M'=0$. The $M=0$ and $J=0$ nonchiral \lags are thus canonical representatives of the super- and sub-critical sectors, for which a non-commutative description analogous to the flat case might be expected \cite{Skagerstam:1989ti}.

\section{BPS Energy Bounds}\label{secBPS}      

The BH spectrum (Fig.1) shows that the critical regions $|M|R=|J|$ are 1/4 BPS supersymmetric configurations, except for $M=0=J$ which is 1/2 BPS. This hints to the existence of BPS energy bounds for the particle moving in AdS$_3$. Indeed, let us consider the energy associated to the chiral \lag \bref{Lchiral2} in the static gauge {($\tau=X^0$)}. When  $MR=J$, the energy bound $E\ge -\mu^+=MR\ge 0$ is saturated by the BPS configuration $X^1_+=X^2_+=0$, but this is not the case when $|J|=-MR$  ($M\leq 0$). Hence, the critical configurations are in correspondence  with the $\frac 14$ BPS supersymmetric states for $M>0$, but not for $M<0$.

\vskip 3mm
\textit{\bf{Acknowledgments}}. We acknowledge Jaume Gomis, Cristi\'an Mart\'{\i}nez and Jorge Russo for discussions. CB partially supported by  Spanish project DPI2011-25649. JG also acknowledges partial financial support from the Dutch Research Organization FOM, from FPA 2010-20807, 2009 SGR502, CPAN, Consolider CSD 2007-0042, and JZ from FONDECYT  1140155.

\end{document}